**The magnetic state of 1212-type ruthenocuprate in magnetocaloric and magnetoresistivity measurements of polycrystalline samples of RuSr$_2$Gd$_{1-x}$Ce$_x$Cu$_2$O$_8$ and Ru$_{1-x}$Sr$_2$GdCu$_2$O$_8$**

Piotr W Klamut [1] and Tomasz Plackowski

Institute of Low Temperature and Structure Research, Polish Academy of Sciences,

P Nr 1410, 50-950 Wrocław 2, Poland

**Abstract**

The magnetic properties of superconducting Ru$_{1-x}$Sr$_2$GdCu$_2$O$_8$ (x=0, 0.02) and non-superconducting RuSr$_2$Gd$_{1-x}$Ce$_x$Cu$_2$O$_8$ (x=0.07, 0.1) were investigated by means of magnetocaloric experiments with complementary magnetoresistivity and ac susceptibility measurements. The isothermal magnetocaloric coefficient $M_T(B)$ assumes positive values in a broad range of temperatures (20K ≤ $T$ ≤ 231 K) and magnetic fields (0 T ≤ $B$ ≤ 13 T), i.e. also in the magnetically ordered state ($T_m$ = 132 K for RuSr$_2$GdCu$_2$O$_8$ and $T_m$=150 K for RuSr$_2$Gd$_{0.93}$Ce$_{0.07}$Cu$_2$O$_8$) which indicates no gain in the system's magnetic entropy with increasing magnetic field. The maximum in the $M_T(B)$ dependence was observed for RuSr$_2$GdCu$_2$O$_8$ in the temperature vicinity of $T_m$, which indicates a ferromagnetic character of the accessed magnetic correlations. No spontaneous ferromagnetic order was revealed as the $M_T$ assumes limiting zero values at zero magnetic field for the whole range of investigated temperatures. Temperature dependencies of the specific heat reveal the magnetic-field-induced positive temperature shift of the anomaly associated with the magnetic transition in the Ru spin system. The $M_T(B)$ dependencies and the magnetoresistivity data suggest that the magnetic system may be inhomogeneous.

arXiv:0810.5558v3

## 1. Introduction

Ruthenocuprates are a family of complex copper oxides crystallizing in a few derivative layered crystal structures, which contain both the $CuO_2$ and $RuO_2$ layers. Several of these compounds were reported as superconducting with $T_C$ up to approximately 50 K bearing close resemblance to other structurally related copper oxide high temperature superconductors (HTSC) [1,2]. The sublattice of Ru ions in ruthenocuprates becomes magnetically ordered at temperatures higher than the superconducting $T_c$, in the range of approximately 70 K - 200 K, depending on the particular structural type [3,4]. This feature makes these compounds distinct from other HTSC and consequently stimulates considerable interest in their properties. Magnetism due to ordering in the Ru sublattice was usually explained by invoking the presence of a ferromagnetic component in the antiferromagnetically aligned sublattice of Ru moments [5-7]. Ferromagnetic properties could certainly add to the fundamental interest, if the emerging phase diagram would involve a coexisting superconducting domain at lower temperatures. Such coexistence indeed seems present for non zero magnetic fields, if one assumes that the crystal lattice has sufficient phase homogeneity for simultaneously supporting both states. This, however, remains as a dispute [8, 9]. Elucidation of the magnetic ground state of the Ru sublattice and its magnetic field dependence proved to be quite complex. Therefore, considerable effort has been addressed in the literature on finding a consistent interpretation of the experimental results provided by several different experimental techniques.

Due to some complexity of the synthesis, comparatively only a few experiments addressed relevant properties in the single crystal and epitaxial thin film form of the ruthenocuprates. With this, we note that the usually reported polycrystalline samples possess a comparatively fine granular structure with the characteristic grain size of approximately 1 μm and are characterized with a comparatively long superconducting magnetic field penetration depth $\lambda(0) \approx 3$ μm [10]. The intrinsic structural inhomogeneity, recently reported for superconducting $RuSr_2GdCu_2O_8$ [8] with high resolution electron microscopy, may be viewed in the context of earlier reports of the phase separation [10]. This further complicates the interpretation of the combined superconducting and magnetic characteristics that have been measured. Manifold properties of the ruthenocuprates were more recently approached in several review texts [10-15]. The systematic approach to the phase diagram for the class, as well as relevant synthesis issues were reviewed in [11].

The $RuSr_2GdCu_2O_8$ is usually indexed in tetragonal symmetry with *P4/mmm* or *P4/mbm* space groups [16,17], although a slight orthorombicity was also reported [18]. In resemblance to the 123-type $GdBa_2Cu_3O_7$ superconductor, the 1212-type structure may be visualized with Ba atoms replaced by Sr, and the Cu chain atoms replaced by Ru coordinated by the octahedra of oxygens, which increases to eight the total number of oxygen atoms per formula. Temperature of the superconducting transition in $RuSr_2GdCu_2O_8$ was found to be synthesis-dependent, with the maximum $T_c$ usually reported at 46 K for the polycrystalline samples and the magnetically established onset $T_c^{on}$=51 K reported in single crystals [19]. By altering the conditions of synthesis also non-superconducting samples of $RuSr_2GdCu_2O_8$ were also found [20,21]. The Ru sublattice in $RuSr_2GdCu_2O_8$ orders magnetically below $T_m \approx$ 131-136 K. The few Kelvin difference in $T_m$ was found for the samples obtained with slightly altered processing conditions, for which the lower $T_m$ usually corresponds to higher superconducting $T_c$ (see [15] and references therein). The assumption of the perfect stoichiometry of the compound and of the Ru ions present in the pentavalent oxidation state, leads to divalent Cu ions, placing the compound below the threshold for superconductivity on the generic phase diagram for HTSC. Partially tetravalent character of the Ru ions was, however, reported in the results of NMR, *i.a.* in [22], and XANES [23]. Then, the average valence of Cu would be expected to increase and may reach into ranges characteristic of the HTSC superconductor.

## 2. Experimental details

In this paper we report on the magnetic properties of polycrystalline samples of $RuSr_2GdCu_2O_8$, $Ru_{0.98}Sr_2GdCu_2O_8$ , $RuSr_2Gd_{0.93}Ce_{0.07}Cu_2O_8$ and $RuSr_2Gd_{0.9}Ce_{0.1}Cu_2O_8$ by means of magnetocalorimetric measurements. We will also discuss the complementary results of the magneto-resistivity and susceptibility measurements.

The samples were prepared in a two step solid state reaction. First, the powders of $RuO_2$ (99.9% purity, preheated in 600°C), $Gd_2O_3$ (99.9% purity), $SrCO_3$ (99.9% purity) and CuO (99.9% purity) were reacted by powder calcination in air at 920°C for 53h with intermediate grinding and then sintering in the controlled flow of Ar gas at 1015 °C for 20h. At this stage, reducing conditions benefited formation of the double perovskite precursor instead of ruthenate strontium perovskite phases, which when formed would subsequently react only with slow kinetics to the final Ru1212-type phase [11]. The X-ray powder diffraction scans confirmed the mixture of $Cu_2O$ and $RuSr_2GdO_6$ phases. Then the samples were sintered at 1060 °C in flowing $O_2$ gas with intermediate grinding and pelletizing, finally cooled at a rate of 1°C/min in the controlled flow of $O_2$ gas and removed from the furnace at 600°C. The samples (except of the sample B of $Ru_{0.98}Sr_2GdCu_2O_8$, which was prepared separately) were synthesized simultaneously with care being taken to provide the same heat treatment conditions for the series. In figure 1 we compared the powder X-ray diffraction spectra for $RuSr_2GdCu_2O_8$ and $Ru_{0.98}Sr_2GdCu_2O_8$ (sample A) in the *2ϑ* range where the diffractive signature of bulk $SrRuO_3$ impurity might be expected.

Compositions of the investigated samples were chosen for accessing the effect of charge doping in the 1212-type structure following conventional routes of crystal chemistry, which are being used in the investigation of many complex copper oxides. Partial substitution of $Ce^{4+}$ ions into the $Gd^{3+}$ site should provide the means for electron doping. Such substituted compounds were indeed found non-superconducting, as would be expected in the case of the lowered Cu valence [24,25]. The Ru deficiency was considered to allow for an effective hole doping in the electronic structure, although we note that such an intuition may well turn misleading for the intrinsically inhomogeneous material. Nevertheless, the superconducting $T_c$ was found higher for nominally deficient $Ru_{0.98}Sr_2GdCu_2O_8$ than in the simultaneously synthesized $RuSr_2GdCu_2O_8$ (see also [15] and references therein). This is in qualitative agreement with the increased $T_c$ reported earlier for $Ru_{0.9}Sr_2GdCu_{2.1}O_8$ in [26]. The ac susceptibility measurements, which show the superconducting transitions in the $Ru_{0.98}Sr_2GdCu_2O_8$ and $RuSr_2GdCu_2O_8$ samples, are presented in figure 2. For both samples, however, the specific heat measurements did not reveal any $T_c$-associated anomaly. The total specific heat of $Ru_{0.98}Sr_2GdCu_2O_8$ in the corresponding temperature range is presented in the inset to figure 2. This agrees with the specific heat data discussed recently in [27] for the superconducting sample of $RuSr_2GdCu_2O_8$, and the discussion of specific heat results provided in [28]. Comparative analysis in [27] estimated the anomaly expected in the electronic part of the specific heat at $T_c$ within one percent range of the measured signal, i.e. below the accuracy of that measurement. Considering the fact that the superconducting phase may form in only constrained volume, it further shifts the limit beyond experimental accuracy.

Temperature dependencies of the dc resistivity were measured by the four contact method with 1mA dc current passed through the bar shaped (approx. 5 mm x 2 mm$^2$) sample. Temperature was changed continuously at 1°C min$^{-1}$ with the error of a single measurement being less than 0.1 K with data collected on heating and cooling. Magnetoresistivity was measured for the magnetic field values in between 0 and 13 Tesla. The same experimental set-up was used for the calorimetric measurements, then with the heat flow calorimeter probe, built for measurements of the specific heat $C_p$ and the isothermal magnetocaloric coefficient $M_T$ [29]. The samples were cut and polished to cuboid size with a foot print of 9 mm$^2$ and mounted on the sensitive heat-flow meter, which contacted the sample with the heat sink. The meter was made of the commercially available, miniature Peltier cells of high thermal conductance. When changing the temperature, the voltage generated on terminals of the Peltier cells is directly proportional to the $C_p$, which was measured on cooling and warming. At a constant temperature and for a steady change of the magnetic field, the generated voltage is proportional to the isothermal magnetocaloric coefficient $M_T$. The $M_T$ was measured at increasing and decreasing magnetic field, with the measurement preceded with precise stabilization of the sample temperature. Temperature dependencies of the specific heat were also measured with the relaxation method using the commercial micro-calorimeter probe in the PPMS measurement platform by Quantum Design Inc. Temperature dependencies of the ac susceptibility were measured with the ac susceptibility probe also in the PPMS platform.

**3. Remarks on synthesis**

Prior to the discussion of the results, which are the merits of the report, we present a few remarks on the investigated samples. The formulas used throughout present their nominal stoichiometries, i.e. reflect the mass proportions of precisely weighted powder substrates. We note that the element selectivity of the sublimation processes, which occur during the synthesis at high temperatures, may effectively alter the final stoichiometry of the sample. For most samples used in this study (except of the sample B of $Ru_{0.98}Sr_2GdCu_2O_8$) the reaction proceeded in the closed glass system in which we noticed a minor film residue aggravating in the cooler part of the apparatus, which was brought there along with the gas which was leaving the heat zone. The EDAX analysis of this residue revealed more of Ru than would be expected based on the sample stoichiometry. We note that, within accuracy of our laboratory powder x-ray diffraction measurements, we could not distinguish between the lattice constants for the $RuSr_2GdCu_2O_8$ and the $Ru_{0.98}Sr_2GdCu_2O_8$ samples. Neither could we quantify the expected element ratio difference with EDAX analysis. Then, the nominal Ru deficiency in one of the samples reflects the direction in which we believe the stoichiometric material could evolve, i.e. for here only a comparatively controlled decrease of the Ru/Cu ratio. Such Ru deficiency, as expressed in the nominal formula, could be accommodated for the finite domain of existence of the 1212-type phase or, alternatively, it should result in the formation of the 1212-type stoichiometric phase accompanied with the other minor phases related by the summation over all of the reaction-available elements. For intrinsically inhomogeneous materials, slight altering of the average stoichiometry should result in a larger amount of these structural inclusions, the formation of which is promoted by the nominal stoichiometry. We note that the recent electron microscopy studies revealed that superconducting $RuSr_2GdCu_2O_8$ is intrinsically inhomogeneous at the nano-scale and its overall structure accommodates the phase inclusions with the diminished Ru/Cu ratio [8]. No data such as that are available for the nominally Ru deficient compounds, which could allow us to investigate the extent of the nano-scale inhomogeneities. The uniformity of the $Ru_{0.98}Sr_2GdCu_2O_8$ sample composition was investigated only at the microscopic scale with conventional scanning electron microscopy. The lower right picture in figure 3 presents the backscatter electrons' scan of the fracture surface area, the same as pictured in the topological scan in the lower left picture. Uniformity of the composition is indicated by the uniform grey colour of the uppermost surfaces seen in the backscatter scan. The synthesis approach applied in this work may be considered complementary to the recent experiments communicated in [30] in which the mass flow control method was used to synthesize samples with slightly different Ru content evident in the $RuSr_2GdCu_2O_8$ structural framework.

The solid state synthesis of here discussed samples proceeded at the temperature of 1060°C, communicated in many reports as the standard conditions of synthesis for most of the ruthenocuprate phases. It means, that the reaction occurs only several kelvins below the melting temperature for the phase. We note that the single phase $RuSr_2GdCu_2O_8$ was also synthesized by the sol-gel method, for which molecular level mixing allowed for a significantly lower temperature of the final reaction [21]. These samples, however, were not superconducting and their considerably smaller grains were invoked for the explanation of the absence of superconductivity. Other possible cause assisting the superconductivity seems to be the alterations of ruthenocuprate stoichiometry [8, 15], which may be promoted during the synthesis in temperature proximity to the phase melting temperature. A potentially related effect was recently communicated for subtle alterations of stoichiometry found induced in a trace melting in the polycrystalline samples of double perovskite $Ru_{1-x}Cu_xSr_2YO_6$ with local intergrowths found to be the superconducting 123-type phase [31]. If such a phase accommodates partial substitution of the Ru ions into the chain Cu sites, it would correspond to the Cu substituted 1212-type. The upper SEM picture in figure 3 presents the selected area in $Ru_{0.98}Sr_2GdCu_2O_8$ (sample B), which was found characteristic by several similar images, which were identified for this sample. The pillar-like region in the central part of the picture has a different morphology than the surrounding fine granular structure, and thus may be suspected of a modified phase character. We note that, in our specific heat data between 175 K and 276 K (not shown), we found the "add-on" feature that could be linked to presence of the $CO_2$ trapped in the sample, possibly during the locally occurring re-solidification process. Analysis of the residual gases present in the commercial $O_2$ gas used in synthesis showed traces of $CO_2$ at 0.08%. We are not aware of the other sources of $CO_2$ that the sample could be exposed to. We note, however, that enthalpy estimation for the area associated with the observed feature corresponds to approximately 0.1 mg of $CO_2$.

## 4. Results and discussion

In the following part we shall discuss the results of our measurements of the specific heat, isothermal magnetocaloric coefficient, magnetic susceptibility and magnetoresistance. The specific heat data of the $RuSr_2GdCu_2O_8$ and $Ru_{0.98}Sr_2GdCu_2O_8$ samples revealed a subtle modification (132 K versus 129 K) of the magnetic transition temperature $T_m$ (see inset to figure 4, also discussed in [15]), which are also observed in a corresponding shift of the ac susceptibility maxima at $T_m$ (not shown). This slightly lower $T_m$ in the sample with nominally deficient Ru sublattice, confirms that its composition was modified compared to the nominally stoichiometric material. For the $RuSr_2Gd_{0.93}Ce_{0.07}Cu_2O_8$ sample the $C_p/T$ bends off at a higher temperature of approximately 150 K (inset to figure 4) and instead of forming a sharp cusp the change is more gradual. We note that there are no characteristic features observed in the specific heat for the Ce-doped sample at temperatures corresponding to $T_m$ of the parent compound. Figure 4 presents the dependencies of $C_p/T$ *vs.* $T$ measured for the $RuSr_2GdCu_2O_8$ at several different magnetic fields in the range between 0 and 13 Tesla. These measurements reveal that the characteristic feature associated with the magnetic transition in the Ru sublattice shifts towards higher temperatures with an increasing magnetic field, as would be considered usual for the ferromagnetic character of dominant magnetic interactions in the corresponding range of magnetic fields.

Since the Cu and Ru ions may formally accept different valence states in the 1212-type structure, the Ce induced electron doping may affect the charge balance between the $CuO_2$ and $RuO_2$ layers. Our specific heat data for the x=0.07 sample show a similar increase of the $T_m$ to that previously communicated in [24,25], and to the effect of the partial substitution of $La^{3+}$ ions into the $Sr^{2+}$ site in $RuSr_{2-x}La_xGdCu_2O_8$ [32,33], all of which seem to be caused by the electron doping. The comparable raise of $T_m$ was, however, communicated also for $RuSr_2Gd_{1-x}La_xCu_2O_8$ [34], i.e. for the isovalent substitution into the Gd site. We should then allow that the modification of the $T_m$ may also be driven structurally, and if allowing for band effects would not only be by differing distance between the Ru ions. Such an example is provided in the properties of the itinerant $Sr_{1-x}Ca_xRuO_3$ (0≤x≤1) where the isovalent substitution induces distortion of the Ru-O-Ru bonds, causing the electronic structure driven modification of magnetic properties. Detailed structural data for the $RuSr_2Gd_{1-x}Ce_xCu_2O_8$ series would allow for further insight. For expected electron doping then lowered Cu valence would result in less conducting Cu-O structural slabs and the sample resistivity should reflect more of the $RuO_2$ layers contribution to scattering. In figure 5 we present the negative magnetoresistance observed for $RuSr_2Gd_{0.93}Ce_{0.07}Cu_2O_8$ and $RuSr_2GdCu_2O_8$. Increase in the conductivity in the magnetic field is more pronounced for the Ce doped sample, in qualitative agreement with the data reported earlier for $RuSr_2Gd_{0.9}Ce_{0.1}Cu_2O_8$ in [25]. The magnetoresistive effect associates with the magnetic transition at $T_m$ at which it is at its largest (arrows in figure 5 show the $T_m$ values as established from the specific heat data). For small magnetic fields the magnetoresistivity becomes positive below $T_m$ (see figure 6), which may be explained with the positive contribution to scattering in the presence of the antiferromagnetic order. Temperature dependencies of the ac susceptibility measured for the $RuSr_2Gd_{0.93}Ce_{0.07}Cu_2O_8$, see figure 7, reveal more complex magnetic properties than only the increased $T_m$ comparing to the parent compound. Both components of the ac susceptibility ($\chi'$, $\chi''$) form maxima at approximately 210 K, which shift to lower temperatures with the *dc* field. We note that only a small *dc* field (see the dependence at $H_{dc}$=500 Oe in figure 7) is required to shift the onset temperature of susceptibility maxima to approximately 170 K and the tendency saturates for higher fields. The ac susceptibility measurement could, in principle, reflect the magnetism of some phase impurity: even the amount of such phases would remain below the detection limit for standard powder x-ray diffraction. However, among other ruthenate phases which may be considered, the highest temperature of magnetic transition belongs to the ferromagnetic $SrRuO_3$ at $T_C$= 163 K, i.e. at a significantly lower temperature than that of the maxima observed in ac susceptibility. The RuSr$_2$GdCu$_2$O$_8$ may be considered charge underdoped in a sense of the HTSC phase diagram and the extra electrons provided with Ce substitution should further that state. Then, in analogy to other complex cuprates, antiferromagnetism of the Cu sublattice could be set at temperatures comparable to 200 K. The negative, field induced shift of the maxima suggests an antiferromagnetic character of the underlying interactions. This tentative interpretation, with a cautionary remark for investigating a possibly inhomogeneous magnetic system, may suggest further element specific measurements like Cu-NQR and Ru-NMR for the RuSr$_2$Gd$_{1-x}$Ce$_x$Cu$_2$O$_8$ series.

In the following part we will analyze the magnetic field dependencies of the isothermal magnetocaloric coefficient $M_T$. These measurements have an advantage of accessing information directly related to the magnetic state of the sample as well as reflect its bulk features. The experiments were performed in a heat flow calorimetric set-up, which is described in detail in [29]. At a constant temperature and steady change of magnetic field conditions, the isothermal magnetocaloric coefficient may be described by the formula:

$$M_T \equiv \frac{dq}{dB} = \frac{-\dot{q}}{\dot{B}} = \frac{-U}{A\dot{B}}$$

where $q$ represents the heat flux from the heat sink to the sample, $B$ is the magnetic induction, $A$ is the sensitivity parameter and $U$ is the voltage generated by the heat flow meter. The negative sign at $q$ preserves consistency with the expression:

$$C_P \equiv \frac{dq}{dT} = \frac{\dot{q}}{\dot{T}} = \frac{U}{A\dot{T}} \ ,$$

which we used for deriving the $C_p$ from data collected at constant magnetic field and changing temperature. Figure 8 presents the magnetic field dependencies of the $M_T$ measured for the RuSr$_2$GdCu$_2$O$_8$ and RuSr$_2$Gd$_{0.9}$Ce$_{0.1}$Cu$_2$O$_8$ samples at temperatures above the magnetic transition temperature $T_m$. Figure 9 shows the $M_T(B)$ dependencies for the same samples at a temperature of 137 K and below the $T_m$. Since two samples were measured simultaneously, the temperature sequences are the same. Change in the magnetic entropy can be accessed by integrating one of the Maxwell relations:

$$\Delta S_{mag}(T, B) = \int_0^B \left(\frac{\partial M}{\partial T}\right)_B dB,$$

and relating the $M_T$ with the magnetization: $M_T = -T\left(\frac{\partial M}{\partial T}\right)_B$ , which leads to:

$$\Delta S_{mag}(T, B) = -\int_0^B \left(\frac{M_T}{T}\right) dB.$$

The positive sign of the isothermal magnetocaloric coefficient, observed for both samples in whole range of the accessed magnetic fields and temperatures (figures 8 and 9), reveals that the system entropy universally decreases with the magnetic field. We note, that for a simple antiferromagnet there is rather an increase in the magnetic entropy, i.e. negative $M_T(B)$ would be expected. Simultaneously, the dependencies differ from the behaviour expected for a simple ferromagnetic system. We note that the presence of spontaneous magnetization would lead to $\left(\frac{\partial M}{\partial T}\right)_{B\to 0} < 0$, i.e. positive values of the $M_T$ in the limit of zero field. Instead, the limiting zero field value of the $M_T$ remains zero for all temperatures below $T_m$, i.e. well into the ordered state (the small shift of approximately 1 J mol$^{-1}$ T$^{-1}$ we attribute to the absolute measurement error). In the magnetically ordered state of the RuSr$_2$GdCu$_2$O$_8$ sample, at temperatures sufficiently close to $T_m$, the $M_T$ raises significantly at low fields and forms the maximum centered at approximately 8 kOe (note the isotherms at T=137 K and at the neighbouring 130.9 K and 134.8 K). In the paramagnetic state, the $M_T$ equals zero at the zero field and approximately obeys the expected linear field dependence. For the characteristic behaviour of the $M_T(B)$ observed in RuSr$_2$GdCu$_2$O$_8$ close to $T_m$, we note that at the phase transition the $M_T$ should follow thermodynamic scaling: $M_T \approx A_m b^{-\omega}$, where $b = \frac{B}{B_c}$ is the reduced magnetic field and $A_m$ is the critical amplitude. It was shown in [35] that the critical exponent $\omega$ should assume comparatively high values, with $\omega$ = 0.4 ± 0.1 found for several universality classes, in particular $\omega \approx 0.39$ for the 3D XY system. Thus, the $M_T$ would diverge rather fast at approaching zero magnetic field. The maximum in $M_T(B)$ observed close to $T_m$ (see figure 8) indicates that the ordering acquires ferromagnetic character at comparatively low magnetic fields. However, within the accuracy of our measurements, no spontaneous ferromagnetic order is revealed at zero field. The isothermal magnetocaloric coefficient measured in analogous experiments for the ferromagnetic UCuP$_2$ [35] and antiferromagnetic UNi$_{0.5}$Sb$_2$ [36] may be referenced for possible comparison. We also conclude that in the investigated range of magnetic fields and temperatures we did not identify the regime in which the system becomes less ordered in result of an application of the magnetic field. The presence of the magnetic domains and randomly aligned crystallites in the polycrystalline specimens of an anisotropic magnetic compound, may also contribute to the rounded character of the observed maximum. The $M_T(B)$ dependences for the RuSr$_2$Gd$_{0.9}$Ce$_{0.1}$Cu$_2$O$_8$ sample (figures 8 and 9) are different in the sense that there is no maximum formed at low magnetic fields at temperatures, which are close to the expected magnetic transition temperature. Note that the specific heat for this sample instead of forming a sharp feature at $T_m$, shows only a gradually changing slope with onset at 150 K (see inset to figure 4). Both indicate a pronounced disorder in the magnetic spin system of RuSr$_2$Gd$_{0.9}$Ce$_{0.1}$Cu$_2$O$_8$. What seems common for all the $M_T(B)$ dependences, also measured below $T_m$, i.e. in the magnetically ordered state, is the lack of saturation of the $M_T(B)$ in the magnetic field. We may conclude that it reflects the significant paramagnetic contribution to the system magnetic entropy, most probably originating in the subsystem of the Gd magnetic moments, which alone remains paramagnetic down to 2.5 K [5]. Due to the degraded sensitivity of our $M_T(B)$ measurements at lower temperatures, we only present the results above 90 K. We do, however, note that for the $RuSr_2GdCu_2O_8$ the $M_T$ at the lowest accessed temperature of 20 K did not change to a negative at the low field as would be expected for a reflecting bulk superconducting phase [37]. Instead, the $M_T(B)$ dependences at 20 K were found similar for both the x=0 and 0.7 samples.

Figure 10 presents the magnetic field dependencies of the magnetoresistivity measured for $Ru_{0.98}Sr_2GdCu_2O_8$ at several temperatures below and above $T_m$. The dependencies become linear in the magnetic field at $T_m$, which becomes the magnetic field induced at higher temperatures. Positive magnetoresistivity emerges at temperatures considerably lower than $T_m$ and only at low magnetic fields, for the higher fields assuming negative with the linear field dependence. Such behaviour may be explained with the use of two components which are simultaneously contributing to the scattering: firstly, the ferromagnetic component which is field-induced in a broad range of temperatures, also well above $T_m$, and the positive antiferromagnetic contribution which becomes dominant only at temperatures considerably lower than $T_m$ and at low magnetic fields. The additional effect of anomalously lowered resistivity observed in temperature vicinity of the onset of the superconducting transition for intermediate fields of approximately 1 Tesla (see the isotherm at 50 K in figure 10) is commented on [2].

Most of the models of the magnetic ordering in the Ru sublattice were formulated for the experimental data collected for the polycrystalline samples, for which it is difficult to account for expected anisotropy of the magnetic system. It is then worth mentioning the recent analysis of the magnetization data for epitaxial thin films of $RuSr_2GdCu_2O_8$, in which the ferromagnetic component of ordering was considered to form cluster type domains out of planar components of the Ru moments, proposed there to couple within $a$-$b$ planes with weak and long range dipolar interactions [38]. The model was essentially meant for reflecting on the experimentally suggested frustrated magnetic ground state and for the presence of the magnetic-field induced ferromagnetism in the $a$-$b$ plane of the investigated films. Since our measurements of the isothermal magnetocaloric coefficient indicate the absence of the spontaneous ferromagnetism, this would match the conclusion formed in [38] with the additional requirement that the antiferromagnetic order would not dominate the system's net magnetic entropy. The positive magnetoresistivity (see figures 6 and 10) is limited to low temperatures and weak magnetic fields, which also suggests the otherwise dominant role of ferromagnetic correlations affecting the electron scattering. The $M_T(B)$ dependencies for $RuSr_2GdCu_2O_8$ also qualitatively agree with the recently proposed A-type antiferromagnetic ground state [39] in which the net ferromagnetism comes only induced with a magnetic field. There, the ferromagnetic easy axis was proposed to lay in the crystallographic $a$-$b$ plane (i.e. along the $RuO_2$ and $CuO_2$ layers) and the spin flop transition was required to transform the system to long range ferromagnetic for a parallel alignment of the $a$-$b$ component layer magnetic moments [39]. It seems that considering different models of the localized Ru magnetic moments the required structure seems to be involving complex anisotropy of the magnetic interactions. On reflection that should come in the observed field-induced positive shift of the temperature of the specific heat anomaly associated with the transition to the magnetically ordered state of Ru for which, by other measurements, an antiferromagnetic order is expected to dominate. When probing thermodynamics of the multi-component magnetic system, there is a possibility of accessing induced magnetic polarization effects, for $RuSr_2GdCu_2O_8$ in the sublattice of large Gd magnetic moments. Our $M_T$ measurements, while indicating the dominance of ferromagnetic type correlations, and not the spontaneous long range ferromagnetism, may then not rule out the antiferromagnetic background in the Ru sublattice. To further comment on the possible multi-component nature of the magnetic system in $RuSr_2GdCu_2O_8$ one should note a different character for the two shown Ru signals in the zero field NMR [39]: the itinerant like for $Ru^{4+}$, and localized for the $Ru^{5+}$ ions. It seems then the contribution of polarized band effects may also be considered when approaching the magnetic-field dependent magnetism of the samples.

In conclusion, the measurements of the isothermal magnetocaloric effect have been communicated for a broad range of temperatures and applied magnetic fields up to 13 Tesla in the magnetically ordered and paramagnetic states of $RuSr_2GdCu_2O_8$ and $RuSr_2Gd_{0.9}Ce_{0.1}Cu_2O_8$ . The results show no gain in the system's magnetic entropy for the investigated range of the magnetic fields and temperatures, indicating a dominant ferromagnetic character of the accessed magnetic interactions. Simultaneously, no spontaneous ferromagnetic order was shown in the $M_T$ data in favor of the magnetic-field-induced net ferromagnetism. The specific heat data reveal that the anomaly associated with the magnetic transition in the Ru spin system shifts to higher temperatures with an increasing field. The $M_T(B)$ and magnetoresistivity in the broad range of temperatures seems to also be consistent with the response of the inhomogeneous magnetic system with a more disorder present for the Ce-doped ruthenocuprate. The superconducting transition temperature for $Ru_{0.98}Sr_2GdCu_2O_8$ is found to be higher compared to that of the nominally stoichiometric sample. However, no bulk superconductivity could be proved in the specific heat data.

**Acknowledgements**

PWK expresses gratitude to Dr Russ Cook of the Materials Science Division of Argonne National Laboratory for the electron microscopy analysis of several samples investigated within NSF Grant DMR-0105398, the SEM pictures of one of which are included in figure 3. The ac susceptibility measurements were performed at the University of Zürich and the author thanks Prof. Hugo Keller at that university for his hospitality. The research was financed by the Polish Ministry of Science and Higher Education research project funding for the years 2007–2010.

---

[1]      Author to whom any correspondence should be addressed. E-mail:P.Klamut@int.pan.wroc.pl

[2]      We associate this anomaly with the negative contribution to magnetoresistivity induced by the superconducting phase. The effect may be approached with the concept of a matching field for the minimum in the vortex phase driven energy dissipation in the superconducting phase, or eventually for a more exotic, since enhanced by the magnetic field, superconducting fluctuations. Note that in the angle dependent *I-V* characteristics of the $RuSr_2GdCu_2O_8$ single crystals, the gradual decrease of the induced voltage was observed for the external field being rotated to parallel to the superconducting *a-b* layers. This was interpreted as indicative of a decreasing number of pancake vortices in the superconducting slabs of the system approximated by the magneto-superconducting multilayered structure [40]. For the field dependent anisotropy of this magnetic system, we might be accessing a similar effect by the field-induced changes in the direction of the magnetic induction vector, which acts on the superconducting layers in the crystallites [41].

arXiv:0810.5588v3

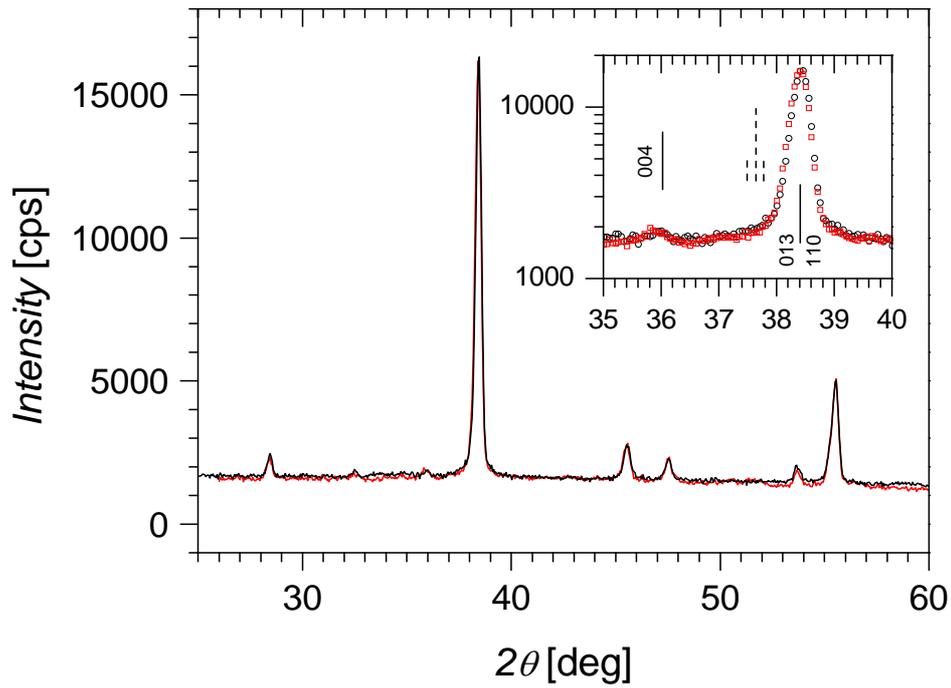

Figure 1. Powder X-ray diffraction spectra of RuSr$_2$GdCu$_2$O$_8$ - (black line, circles in the inset), and Ru$_{0.98}$Sr$_2$GdCu$_2$O$_8$ (sample A, red line, squares in the inset). Inset shows the scans at expanded scale in the range where the main diffraction maxima of the SrRuO$_3$ phase are expected. The vertical lines mark positions of the 004, 013 and 110 maxima for the RuSr$_2$GdCu$_2$O$_8$ derived in the tetragonal *P4/mmm* space group. The dashed vertical lines from left to right mark positions of the 200, 112 and 020 maxima for SrRuO$_3$ in the orthorhombic *Pbnm* space group. *Co Kα* radiation, T=293 K.

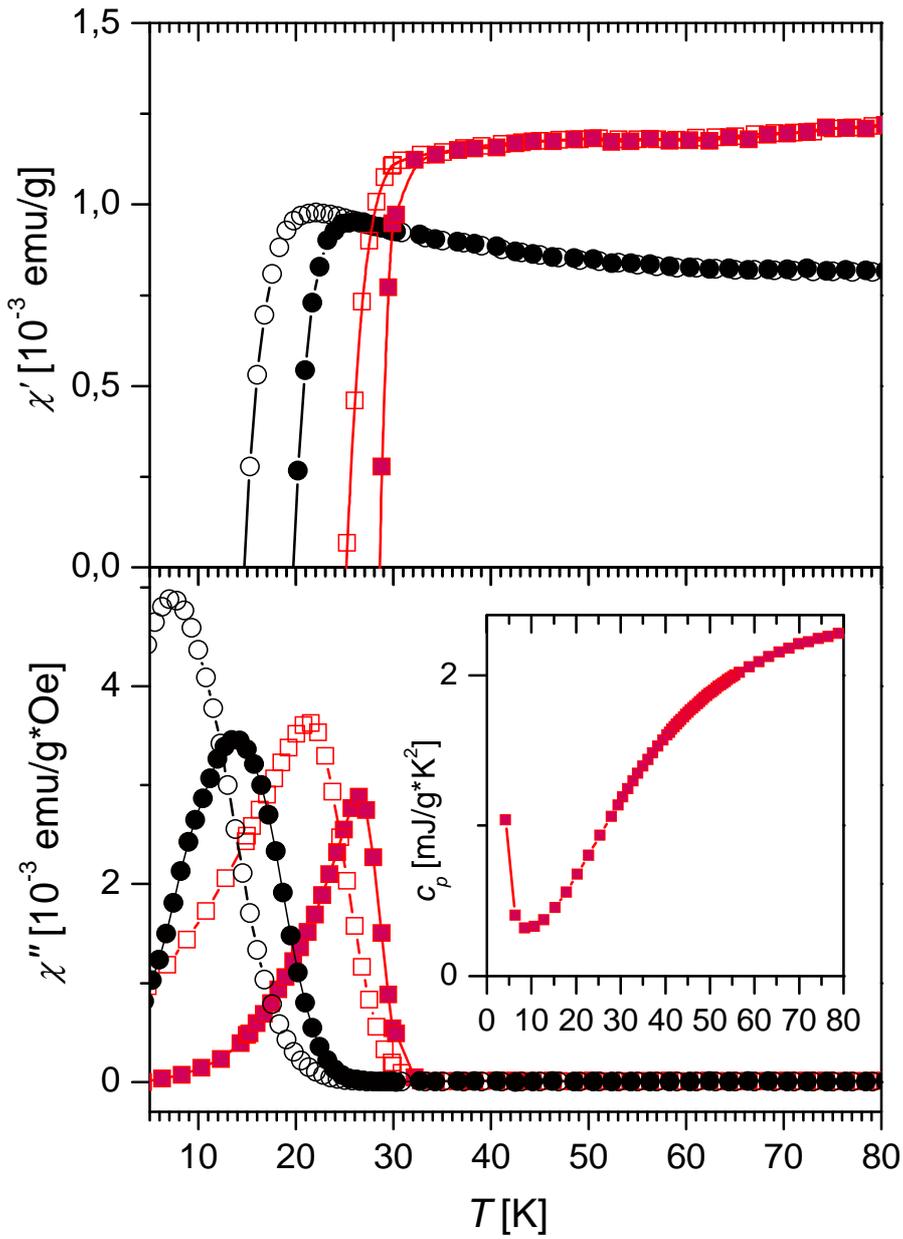

Figure 2. Temperature dependencies of the real and imaginary components of *ac* susceptibility in a vicinity of the superconducting transitions in Ru$_{0.98}$Sr$_2$GdCu$_2$O$_8$ (sample A, squares, red in colour), and RuSr$_2$GdCu$_2$O$_8$ (circles, black in colour). Open symbols correspond to the ac field $H_{ac}$=1 Oe, closed symbols to $H_{ac}$=0.1 Oe, $f$=1 kHz. Inset presents C$_p$/$T$ *vs. T* for Ru$_{0.98}$Sr$_2$GdCu$_2$O$_8$ (sample A) in the corresponding range of temperatures.

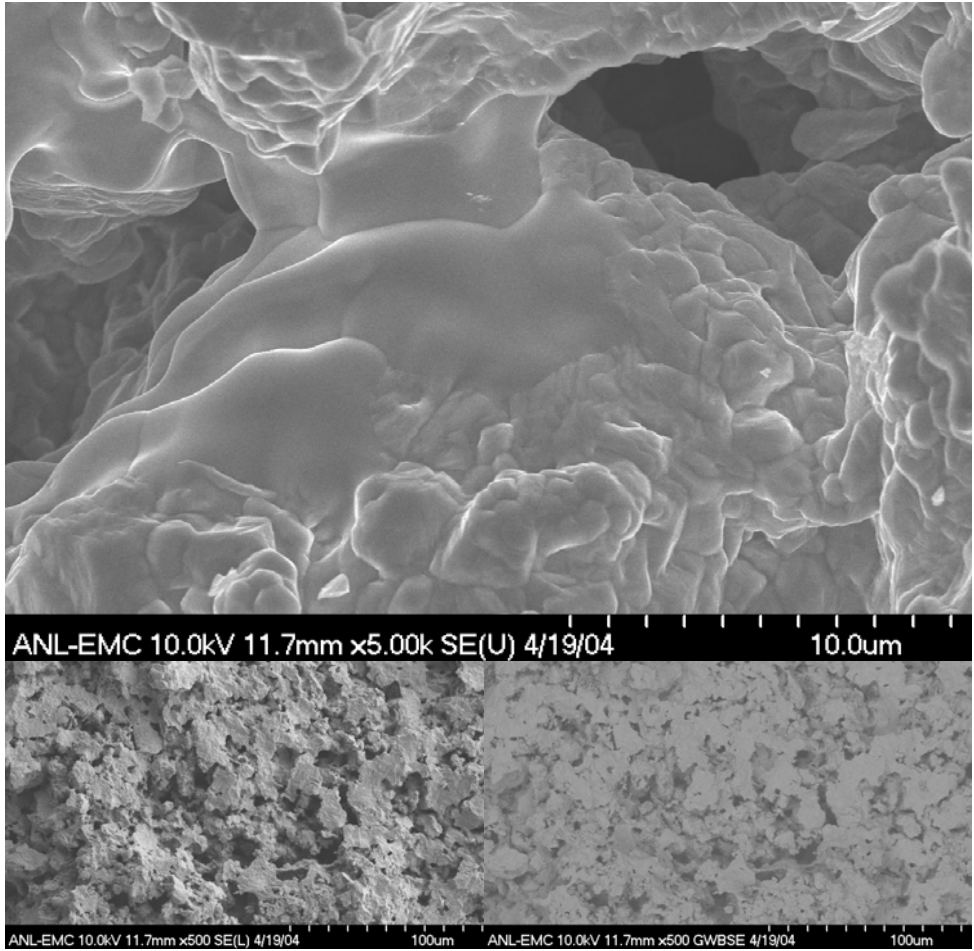

ANL-EMC 10.0kV 11.7mm x5.00k SE(U) 4/19/04          10.0um

ANL-EMC 10.0kV 11.7mm x500 SE(L) 4/19/04     100um     ANL-EMC 10.0kV 11.7mm x500 GWBSE 4/19/04     100um

arXiv:0810.5588v3

Figure 3. SEM pictures of the $Ru_{0.98}Sr_2GdCu_2O_8$ (sample B). The upper picture shows selected
characteristic area of the sample. Lower pictures are the large area images of fracture
surface: left presents the topographical view of the area which is presented in the right
picture with the backscatter detector in composition mode – there no greyscale contrast at
the uppermost surfaces indicates uniformity of sample composition.

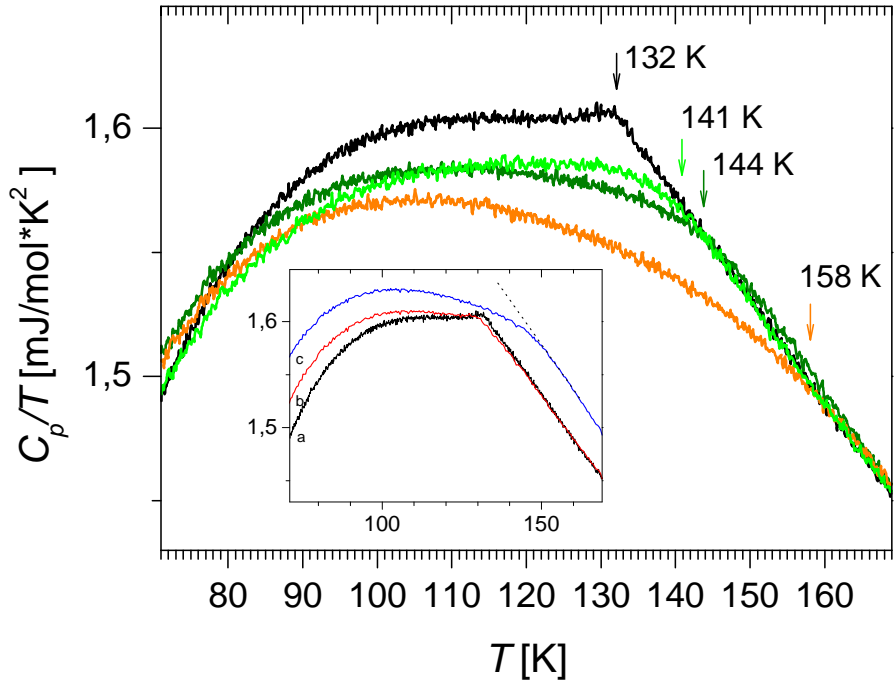

Figure 4. $C_p/T$ vs. $T$ in a vicinity of $T_m$ for RuSr$_2$GdCu$_2$O$_8$ (sample A) measured at different magnetic fields: 0 T (black in colour), 1 T (green in colour) ,3 T (dark green in colour) and 13 T (orange in colour). Arrows show the magnetic transition temperatures, which increase with the field. Inset shows the $C_p/T$ vs. $T$ for: (a) RuSr$_2$GdCu$_2$O$_8$ (black in colour), (b) Ru$_{0.98}$Sr$_2$GdCu$_2$O$_8$ (red in colour) and (c) RuSr$_2$Gd$_{0.93}$Ce$_{0.07}$Cu$_2$O$_8$ (blue in colour), also in [15].

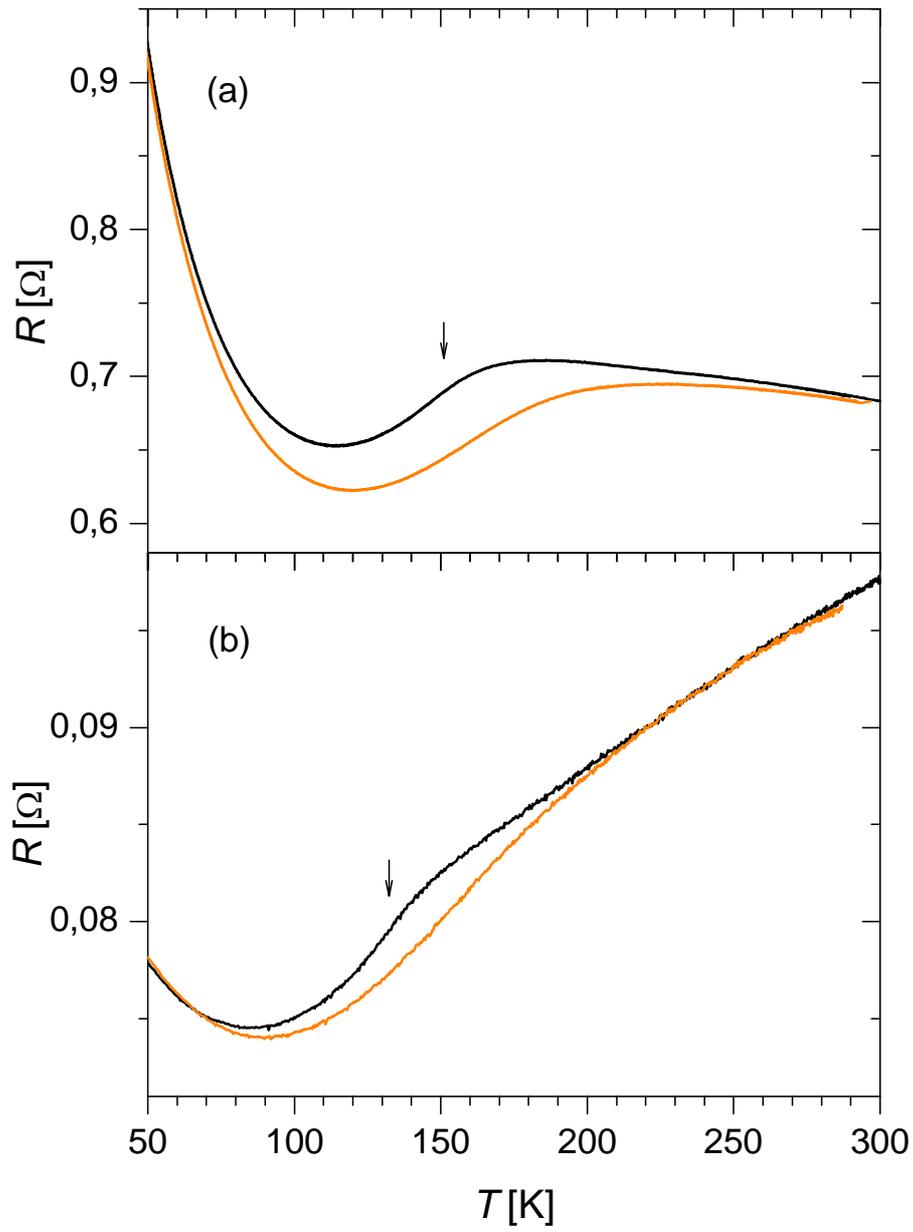

Figure 5. Negative magnetoresistance for samples:(a) $RuSr_2Gd_{0.93}Ce_{0.07}Cu_2O_8$ and (b) $RuSr_2GdCu_2O_8$.
Dependencies shown are measured at: 0 T (black in colour, higher resistance values) and 13
Tesla (orange in colour, lower resistance values). Arrows show the temperatures of
magnetic transitions as established in the specific heat measurements.

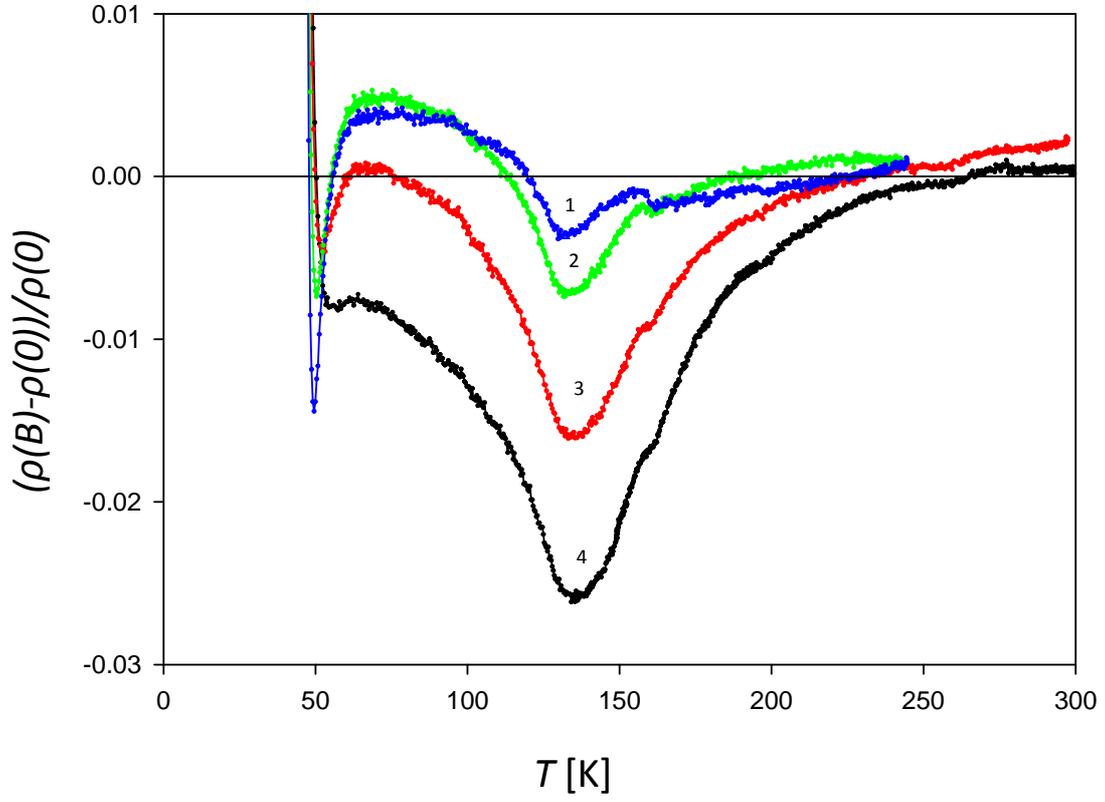

Figure 6. Temperature dependences of magnetoresistivity for sample B of $Ru_{0.98}Sr_2GdCu_2O_8$. Different curves correspond to different values of magnetic field: 1) 1 T (blue in colour), 2) 3 T (green in colour), 3) 7 T (red in colour) and 4) 13 T (black in colour).

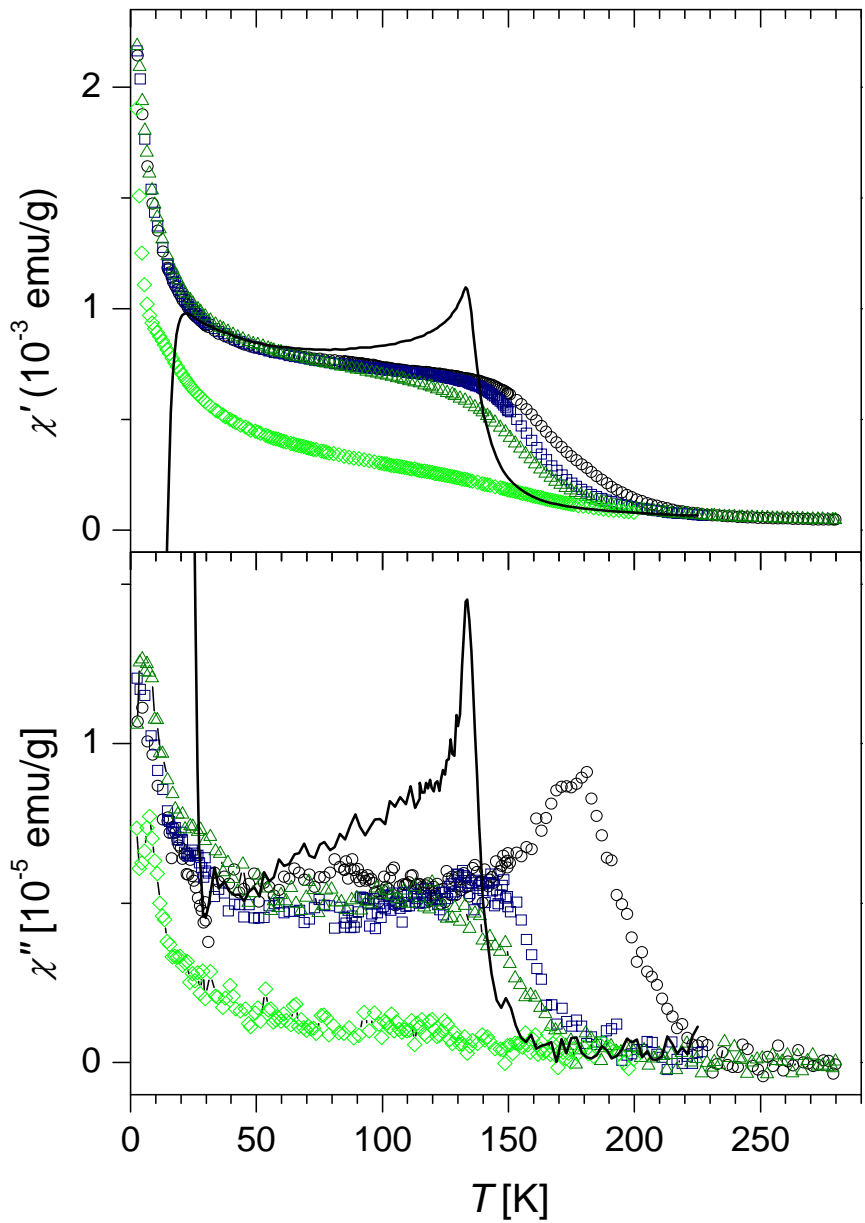

Figure 7. Temperature dependencies of the real and imaginary components of ac susceptibility for RuSr$_2$Gd$_{0.93}$Ce$_{0.07}$Cu$_2$O$_8$ at different values of the dc field: $H_{dc}$=0 Oe (circles, black in colour), 500 Oe (squares, blue in colour), 1 kOe (triangles, dark green in colour) and 10k Oe (rhombs, green in colour). Solid line shows the data for the RuSr$_2$GdCu$_2$O$_8$ sample at H$_{dc}$=0 Oe. $H_{ac}$=1 Oe, $f$=1 kHz.

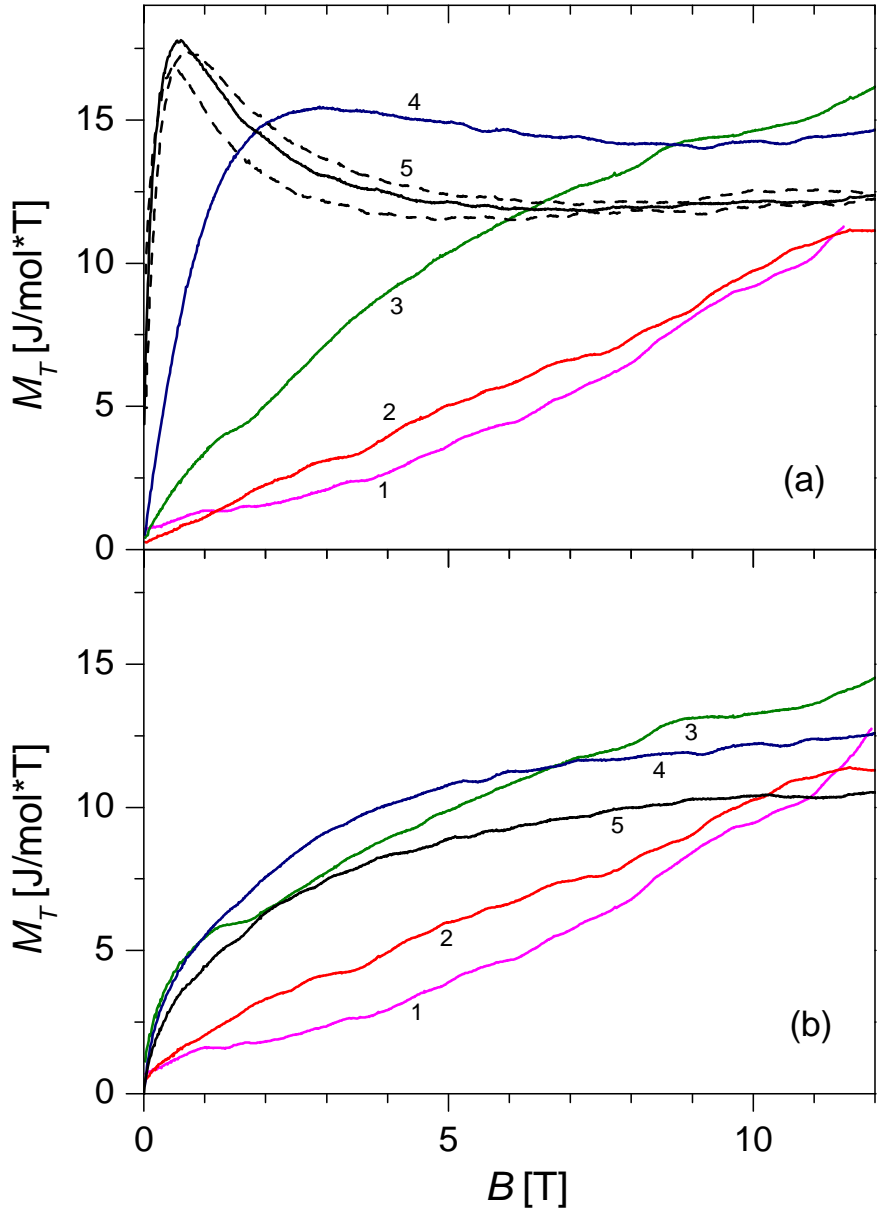

Figure 8. Magnetic field dependencies of isothermal magnetocaloric coefficient for: (a) RuSr$_2$GdCu$_2$O$_8$ and (b) RuSr$_2$Gd$_{0.9}$Ce$_{0.1}$Cu$_2$O$_8$ measured at different temperatures above $T_m$ : 1) T=230.9 K (magenta in colour), 2) 200.6 K (red in colour), 3) 170.3 K (green in colour), 4) 150.0 K (blue in colour), 5) 137.0 K (black in colour). Dashed line dependencies for RuSr$_2$GdCu$_2$O$_8$ correspond to 138.9 K (upper) and 134.8 K (lower), which neighbour the dependence at 137 K, for which highest maximum at low fields was found.

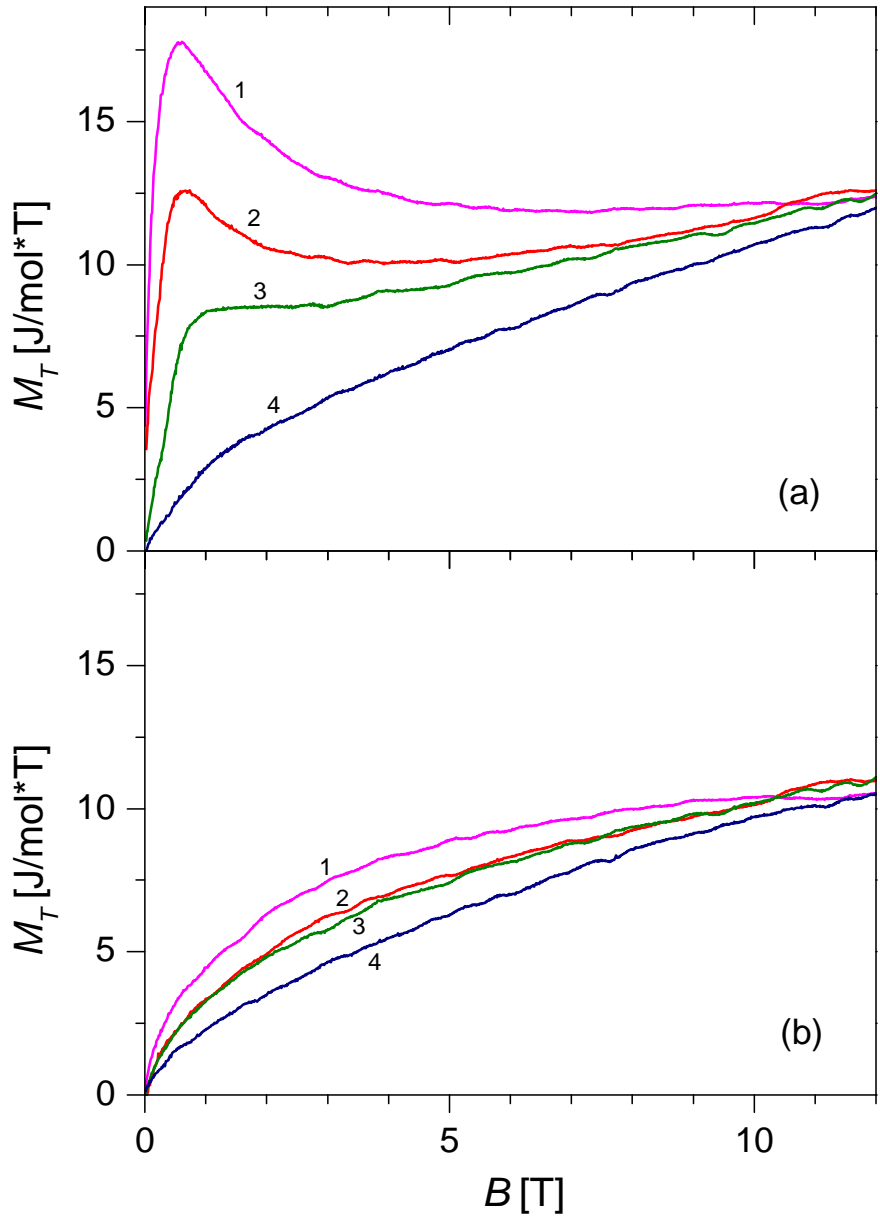

Figure 9. Magnetic field dependencies of isothermal magnetocaloric coefficient for: (a) RuSr$_2$GdCu$_2$O$_8$ and (b) RuSr$_2$Gd$_{0.9}$Ce$_{0.1}$Cu$_2$O$_8$ measured at different temperatures at 137 K and below $T_m$: 1) T=137.0 K (magenta in colour), 2) 130.9 K (red in colour), 3) 124.8 K (green in colour) and 4) 89.8 K (blue in colour).

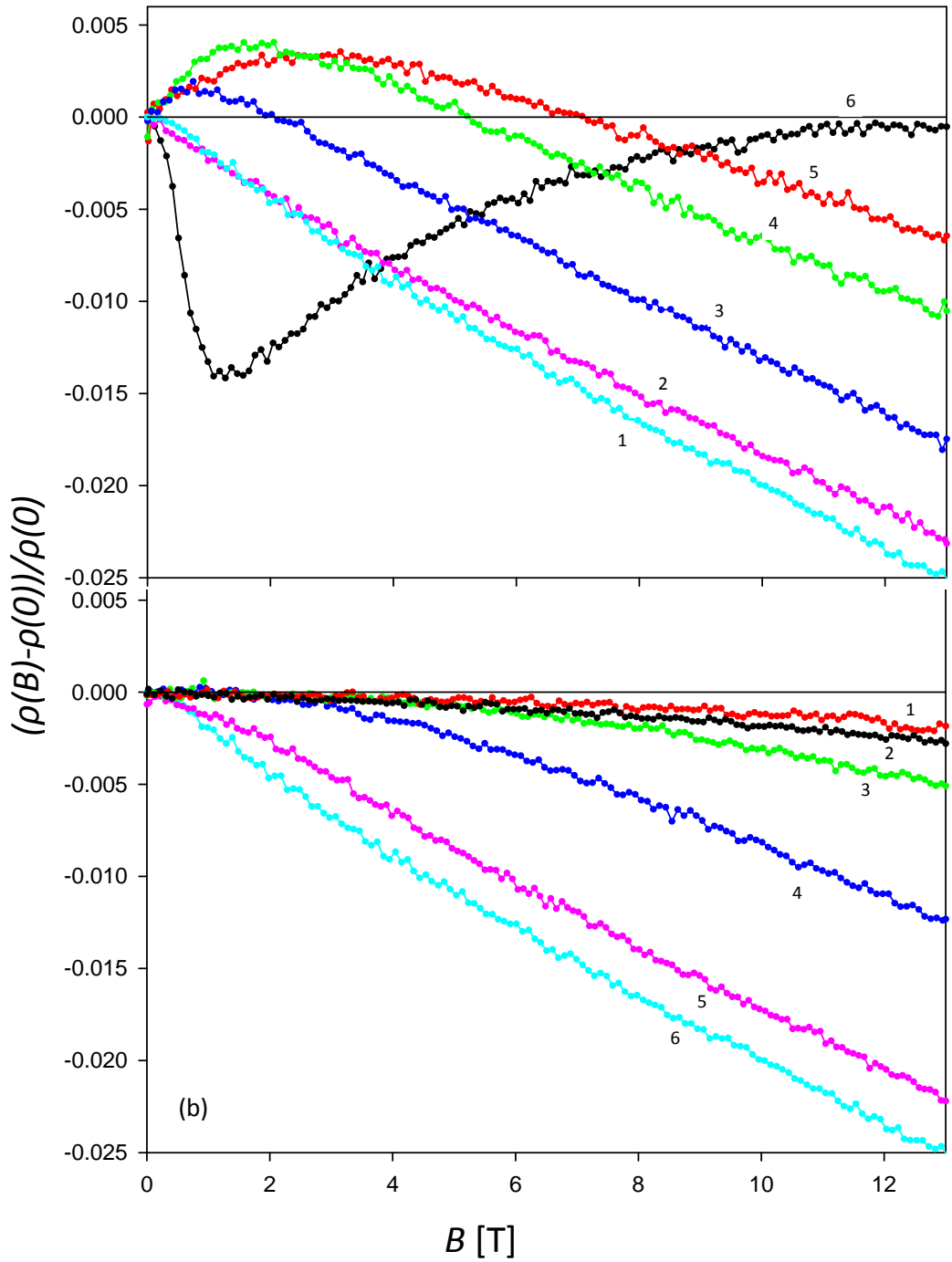

Figure 10. Magnetic field dependencies of magnetoresistivity for Ru$_{0.98}$Sr$_2$GdCu$_2$O$_8$ (sample B) at different temperatures below (a) and above (b) $T_m$. Temperatures for (a) are: 1) 140 K (blue in colour), 2) 130 K (purple in colour), 3) 120 K (dark blue in colour), 4) 100 K (green in colour), 5) 60 K (red in colour) and 6) 50 K (black in colour). Temperatures for (b): 1) 300 K (red in colour), 2) 250 K (black in colour), 3) 200 K (green in colour), 4) 170 K (dark blue in colour), 5) 150 K (purple in colour), 6) 140 K (blue in colour).